\documentclass[aps, prd, amsmath, floats, floatfix, onecolumn, nofootinbib]{revtex4-2}
\pdfoutput=1
\usepackage{graphicx}
\usepackage{color}
\usepackage{latexsym}
\usepackage{amsmath}
\usepackage[makeroom]{cancel}
\usepackage[makeroom]{cancel}

\usepackage{amssymb}

\usepackage[hidelinks]{hyperref}

\newcommand{\be}{\begin{eqnarray}}
	\newcommand{\ee}{\end{eqnarray}}

\newcommand{\bea}{\begin{eqnarray}}
	\newcommand{\eea}{\end{eqnarray}}

\begin{document}

	\title{Conformalons: a new class of black hole mimickers} 
	
	\author{Leonardo Modesto}
	\email[]{lmodesto1905@icloud.com}
	\affiliation{Dipartimento di Fisica, Universit\`a di Cagliari, Cittadella Universitaria, I-09042 Monserrato, Italy}
	\affiliation{INFN, Sezione di Cagliari, Cittadella Universitaria, I-09042 Monserrato, Italy}
	
	\author{Ali Akil}
	\email[]{aliakil@hku.hk}
	\affiliation{Quantum Information and Computational Initiative (QICI), Department of Computer Science, The University of Hong Kong, Hong Kong, China}

	\author{Cosimo~Bambi}
	\email[]{bambi@fudan.edu.cn}
	\affiliation{Center for Astronomy and Astrophysics, Center for Field Theory and Particle Physics, and Department of Physics,
Fudan University, Shanghai 200438, China}
	\affiliation{School of Natural Sciences and Humanities, New Uzbekistan University, Tashkent 100007, Uzbekistan}

	\begin{abstract}
		In any conformally invariant gravitational theory, the space of exact solutions is greatly enlarged. Therefore, we cannot exclude the Weyl's conformal invariance to be spontaneously broken to spherically symmetric vacuum solutions that exclude the spacetime region inside the black hole's event horizon from our Universe. We baptize these solutions {\it conformalons}. It turns out that for all such spacetimes nothing can reach the Schwarzschild event horizon in a finite amount of proper time for conformally coupled ``massive'' particles, or finite values of the affine parameter for massless particles. Therefore, for such vacuum solutions the surface $r = 2 M$ becomes an asymptotic region of the Universe. As a general feature, all conformalons show a gravitational blueshift instead of a gravitational redshift at the unattainable Schwarzschild event horizon, hence avoiding the Trans-Planckian problem in the Hawking evaporation process. Unexpectedly, the Hawking's temperature of the conformalons turns out to be negative and the thermodynamic interpretation leads us to speculate about a possible maximum energy state of the Hawking radiation. Contrary to the Schwarzschild spacetime, for the conformalons the gravitational collapse consists of matter that falls down forever towards the Schwarzschild horizon without ever reaching it. Hence, the annihilation process between the matter and Hawking's negative energy particles takes place outside the surface at $r = 2 M$. Finally, the information is not lost in the whole process of singularity-free collapse and evaporation.
	\end{abstract}

	\maketitle
	
	\section{Introduction}

	In Einstein's conformal gravity (ECG), which we will shortly remind the reader of, as well as in any conformally invariant theory solved by Ricci-flat spacetimes, a general solution takes the following form, 
	\be
	d \hat{s}^{* 2} = \hat{g}^*_{ \mu\nu}(x) dx^\mu d x^\nu = 
	S(x) \hat{g}_{\mu\nu}(x) dx^\mu d x^\nu , 
	\label{Rescaling} 
	\ee
	where the rescaling factor $S(r)$ is the vacuum selected through the spontaneous symmetry breaking of conformal invariance. Namely, a geodesically complete vacuum is naturally selected without explicitly breaking the conformal symmetry.  
	Indeed, in the conformal phase there is no singularity issue as long as the singular spacetimes are conformal to those that are singularity free. On the other hand, the vacuum in the spontaneously broken phase is consistently selected in order to interpolate with the  singularity-free spacetime in the conformal phase. 
	
	From now on, we will focus on ECG, but most of the results can be extended to Weyl conformal gravity and all the other theories invariant under the following conformal symmetry\footnote{See, for example, the theories proposed in~\cite{Krasnikov,kuzmin,modesto,review,modestoLeslaw}, which are consistent with quantum super-renormalizability or finiteness, linear and non linear stability \cite{StabilityMinkAO, StabilityRicciAO}, and perturbative unitarity \cite{cutkosky,Briscese:2018oyx}. In particular, the theory in \cite{modesto} in odd dimension and the generalization in even dimension \cite{modestoLeslaw} are both finite at quantum level, a property that turns out to be crucial in order to support the main statement in this paper. Indeed, here we assume that conformal symmetry is broken spontaneously and not by the quantum Weyl anomaly. Therefore, the gravitational theory has to be finite at quantum level \cite{Universally,FiniteGaugeTheory,Modesto:2021ief,Modesto:2021okr,Calcagni:2014vxa,Giaccari:2016kzy}.}
	\be
	&& \hat{g}_{\mu\nu}^\prime(x) = \Omega^2(x) \hat{g}_{\mu\nu}(x) \, , \nonumber \\
	&& \phi^\prime(x) = \Omega^{-1}(x) \phi(x)  \, ,
	\label{Winv}
	\ee
	where $\phi$ is the dilaton field that guarantees the Weyl invariance of the action. 
	
	The action for ECG in a $D-$dimensional spacetime is obtained replacing 
	\be
	g_{\mu\nu} = \left( \phi^2 \kappa_D^2 \right)^{\frac{2}{D-2}} \hat{g}_{\mu\nu} \, ,
	\label{phighat}
	\ee
	in the Einstein-Hilbert gravitational action 
	\be
	S_{\rm EH} =  \frac{2}{ \kappa_D^2} \int\!d^Dx\sqrt{|g|} R(g)\, , 
	\label{EH}
	\ee
	where $2/\kappa_D^2=1/16 \pi$ for $G = 1$. The resulting Lagrangian reads: 
	\be
	\mathcal{L}
	=   
	2  \, \sqrt{ |\hat{g}| } 
	\left[   \phi^2 \hat{R}(\hat{g}) +
	\frac{4(D-1)}{ D-2} \hat{g}^{\mu\nu} (\partial_\mu \phi) (\partial_\nu \phi) \right] .
	\label{CEHG}
	\ee
	If the conformal symmetry is spontaneously broken to a constant value for the dilaton field, namely $\phi = 1/\kappa_D$, the action in (\ref{CEHG}) turns into the Einstein-Hilbert one in (\ref{EH}). 
	Notice that there is no propagating degree of freedom related to the dilaton field because it can be always gauge fixed to zero making use of the conformal invariance. 
	
	In several papers it has been proved that the singularity issue can be successfully addressed in any conformally invariant theory having Ricci flat spacetimes as exact solutions 
	\cite{Bambi:2017ott, Bambi:2016yne, Modesto:2016max, Modesto:2018def, Zhou:2019hqk, Zhang:2018qdk, Zhou:2018bxk, Chakrabarty:2017ysw, Jusufi:2019caq}. 
	Indeed, by proper rescaling, we can obtain regular black hole solutions starting from the Schwarzschild one. In this case the conformal symmetry is spontaneously broken to a non trivial profile for the dilaton instead of the constant one discussed above, in which conformal gravity turns into Einstein's gravity. For all such singularity-free spacetimes, it turns out that nothing can reach the singularity in a finite proper time (massive particles) or for a finite value of the affine parameter (photons or massless particles). Therefore, the spacetime point $r=0$ is outside of our Universe. 
	Conformal symmetry can also play a crucial role in explaining the galactic rotation curves as explicitly shown in \cite{Li:2019ksm}. 
	
	In this paper, motivated by the above results, we fully remove the whole black hole interior from our Universe by the means of a conformal rescaling that is singular at the event horizon.

	For the sake of simplicity, in the rest of the paper we will assume to be in $D=4$, but it is straightforward to generalize all the results in this paper to any spacetime dimension $D$.
	
	\section{Conformalons}
	
	As a consequence of the Weyl invariance (\ref{Winv}), in ECG the Schwarzschild solution and all its rescalings are exact solutions of the theory (see equation (\ref{Rescaling})). Therefore, we hereby assume the Weyl conformal symmetry to be spontaneously broken to the following metric that exactly solve the equations of motion for the theory (\ref{CEHG}), 
	\be
	&& d \hat{s}^{* 2} = S(r) \left[ - \left( 1 - \frac{2 M}{r} \right) dt^2 + \frac{dr^2}{1 - \frac{2 M}{r}} + r^2 d \Omega^{(2)} \right] \, , \nonumber \\
	&& \phi^*(r) = S^{-1/2}(r) \kappa^{-1}_4 \, , 
	\label{UBHmetric}
	\ee
	where for the rescaling $S(r)$, we consider multiple choices, namely,
	\be
	&& S(r) \equiv  \frac{r}{r- 2 M}  \label{incomplete} \, , \\
	&& S(r) \equiv \left( \frac{r}{r- 2 M} \right)^2  \label{minimal} \, ,\\
	&& S(r) \equiv \left( \frac{r^{20}}{r^{20}- (2 M)^{20}} \right)^2 \, , \label{red-blue2} \\
	&& S(r) \equiv \left( \frac{r^{n}}{r^{n}- (2 M)^{n}} \right)^2 \, , \quad n \in \mathbb{N} \, ,  \label{red-bluen} \\
	&& S(r) \equiv \left( \frac{r^{n}}{r^{n}- (2 M)^{n}} \right)^m  \, , \quad n, \, m \in \mathbb{N}\label{red-bluenm} \,\,  , \, m > 2\, . 
	\ee
	We name these solutions {\em conformalons} because, as it will be clear in the next section, the spacetime is  defined only for $r> 2 M$ (similar to gravitational instantons), while the interior of the black hole is an unreachable region for any massive or massless particle. In other words, conformalons are asymptotically unattainable regions of the Universe.

	Since the rescaling excludes the spacetime region inside the event horizon [$S(r)$ is only defined for $r>2 M$], as a first check of the regularity of the spacetime we consider the Kretschmann invariant for $r \geqslant 2 M$. 
	The Schwarzschild spacetime $\hat{g}_{\mu\nu}$ is Ricci flat, hence, as it is well known, before the rescaling the Kretschmann scalar is:
	\be
	\hat{K} := \hat{R}_{\alpha \beta \gamma \delta} \hat{R}^{\alpha \beta \gamma \delta} = \hat{C}_{\alpha \beta \gamma \delta} \hat{C}^{\alpha \beta \gamma \delta} = \frac{ 48 M^2}{r^6} \, , 
	\label{Kre}
	\ee
	where in the last equality we used that $\hat{R}_{\alpha \beta}=0$ and introduced the Weyl tensor $\hat{C}_{\alpha \beta \gamma \delta}$. Under the Weyl rescaling (\ref{Winv}) the Weyl tensor is invariant, i.e.  
	\be
	\hat{C}^{\prime \alpha}\,_{ \beta \gamma \delta} = \hat{C}^{\alpha}\,_{ \beta \gamma \delta}\, , 
	\ee
	while the Kretschmann scalar (\ref{Kre}) turns into 
	\be
	\hat{K}^\prime = \hat{C}^{\prime \alpha}\,_{\beta \gamma \delta} \, \hat{C}^{\prime \mu}\,_{\nu\rho\sigma} \,  \hat{g}^\prime_{\alpha \mu} \, \hat{g}^{\prime \beta \nu} \, \hat{g}^{\prime \gamma \rho} \, \hat{g}^{\prime \delta \sigma} 
	= \hat{C}^{\alpha}\,_{\beta \gamma \delta} \, \hat{C}^{\mu}\,_{\nu\rho\sigma} \,  \hat{g}_{\alpha \mu} \, \hat{g}^{\beta \nu} \, \hat{g}^{\gamma \rho} \,  \hat{g}^{\delta \sigma} S(r) \, S^{-1}(r) \, S^{-1}(r)\,  S^{-1}(r) = \frac{\hat{K}}{S^2(r)}  \,  ,
	\label{K}
	\ee
	where the label $^\prime$ is actually $^*$ for the metric in (\ref{UBHmetric}). 
	Therefore, for the metric (\ref{UBHmetric}) and (for example) the rescaling (\ref{minimal}) 
	\be
	\hat{K}^\prime = \hat{K}^* = \frac{\hat{K}}{S^2(r)} = \frac{ 48 M^2}{r^6} \, \left( \frac{r - 2 M}{r} \right)^4  , 
	\ee
	which is zero for $r = 2 M$. Therefore, the Schwarzschild event horizon becomes an asymptotic region of zero curvature (the reason of using here the terminology ``asymptotic region'' will be clear in the next section when we will study the geodesic completeness). Certainly, it is not sufficient to look at the Kretschmann invariant to claim the regularity of the spacetime, but it is a useful tool in order to determine the right rescaling consistently with the spacetime geodesic completeness. It deserves to be noticed that $\hat{K}^\prime$ is zero at  $r = 2 M$ even for the rescaling (\ref{incomplete}), but as we will see later the spacetime is not geodesically complete.

	\section{Geodesic completeness}
	
	In oder to prove the geodesic completeness of (\ref{UBHmetric}) and to understand the global structure of the Universe,
	we can study the motion of conformally coupled particles and of massless particles in the metric (\ref{UBHmetric}). It will turn out that for the rescalings (\ref{minimal}) - (\ref{red-bluenm}) nothing is able to reach the Schwarzschild event horizon at $r = 2 M$ in a finite amount of proper time, or for a finite value of the affine parameter in the case of massless particles. We also could study the propagation of massive particles in the geometry (\ref{UBHmetric}), but 
	such analysis is not very interesting and purely academic since we do not want to explicitly break the Weyl conformal symmetry.  
	
	For the sake of simplicity, from now on we make the following redefinitions: 
	\be
	\hat{g}^*_{\mu\nu} \rightarrow \hat{g}_{\mu\nu} \, , \quad 
	\phi^* \rightarrow \phi \, . 
	\ee

	\subsection{Geodesics for conformally coupled particles}

	We begin studying the geodesic completeness of the spacetime (\ref{UBHmetric}) considering a conformally coupled particle whose action reads: 
	\be
	S_{\rm cp} = - \int \sqrt{ - f^2 \phi^2 \hat{g}_{\mu\nu} d x^\mu d x^\nu} 
	=  - \int \sqrt{ - f^2 \phi^2 \hat{g}_{\mu\nu} \frac{d x^\mu}{d \lambda} \frac{d x^\nu}{d \lambda} } 
	\, d \lambda \, , 
	\label{Spc}
	\ee
	where $f$ is a positive constant coupling strength, $\lambda$ is a parameter, and $x^\mu(\lambda)$ is the trajectory of the particle. Notice that in the unitary gauge $\phi = \kappa_4^{-1}$ the action~(\ref{Spc}) turns into the usual one for a particle with mass $m = f \kappa_4^{-1}$ ($f>0$). The Lagrangian reads 
	\be
	L_{\rm cp} = - \sqrt{ - f^2 \phi^2 \hat{g}_{\mu\nu} \dot{x}^\mu \dot{x}^\nu } \, , 
	\ee
	and the translation invariance in the time-like coordinate $t$ implies 
	\be
	\frac{\partial L_{\rm cp}}{\partial \dot{t} } = - \frac{f^2 \phi^2 \hat{g}_{tt} \dot{t}}{L_{\rm cp}} = {\rm const.} = - E \quad \Longrightarrow \quad \dot{t} = \frac{L_{\rm cp} E }{f^2 \phi^2 \hat{g}_{tt}} , 
	\label{ConstE}
	\ee
	Since we are interested in evaluating the proper time that the particle takes to reach the Schwarzschild event horizon located at $r= 2 M$, we must choose the proper time gauge, namely $\lambda = \tau$. In this case, $E$ is the energy of the test-particle and
	\be
	\frac{d \hat{s}^2}{d \tau^2} = - 1 
	\quad \Longrightarrow \quad  L_{\rm cp}= - f \phi 
	\quad \Longrightarrow \quad  \dot{t} = - \frac{ E }{f \phi \, \hat{g}_{tt}} \, .
	\label{PTG}
	\ee
	Replacing~(\ref{ConstE}) in $\hat{g}_{\mu\nu} \dot{x}^\mu \dot{x}^\nu = - 1$ and using the solution of the EOM for $\phi$, namely $\phi = S^{-1/2} \kappa^{-1}_4$, we end up with 
	\be
	S(r)^2 \dot{r}^2 + S(r) \left( 1 - \frac{2 M}{r} \right) - \frac{E^2 \kappa_4^2}{f^2} S(r) = 0 \, , 
	\ee
	or 
	\be
	S(r) \dot{r}^2 + \left( 1 - \frac{2 M}{r} \right) - \frac{E^2 \kappa_4^2}{f^2}  = 0 .
	\ee
	Introducing the definition $E^2 \kappa_4^2/ f^2 \equiv e^2 > 0$ we finally get:
	\be
	S(r) \dot{r}^2 = e^2 -1 + \frac{2 M}{r}   .
	\label{finalGE}
	\ee
	For $r\gtrsim 2 M$, the latter equation (\ref{finalGE}) simplifies to
	\be
	S(r) \dot{r}^2 \sim e^2   \quad \Longrightarrow \quad \sqrt{S(r)} | \dot{r} | \sim |e| \, .
	\label{finalGE2}
	\ee
	Hence, for the rescaling (\ref{minimal}) and $r \approx 2 M$, the above equation can be easily integrated, 
	\be
	\frac{2 M}{r - 2 M } | \dot{r} | \sim |e| \quad \Longrightarrow \quad 
	\frac{2 M}{| e |} \log (r - 2 M) = - \tau + {\rm const.} 
	\, .
	\label{finalGE3}
	\ee
	Therefore, the test particle cannot reach the surface $r = 2 M$ in a finite proper time. 
	
	Let us now integrate exactly (\ref{finalGE}). 
	Since the rescaling is singular at $r = 2 M$, the radial geodesic equation (\ref{finalGE}) is defined for $r>2 M$ and integrating (\ref{finalGE}) we find the following proper time, 
	\be
	\int \sqrt{\frac{S(r)}{e^2 -1 + \frac{2 M}{r} }} \, dr = - \tau + {\rm const.}\, .
	\label{integral}
	\ee
	The solutions of (\ref{integral}) for the rescalings (\ref{incomplete}) and (\ref{minimal}), assuming $e=1$, are, respectively, 
	\be
	&& \tau = \frac{2}{3} \left[ - \frac{r + 2 r_g}{\sqrt{\frac{r_g}{r- r_g}}} +  \frac{r_0 + 2 r_g}{\sqrt{\frac{r_g}{r_0- r_g}}}
	\right] \, , \\
	&& \tau = \frac{2}{3} \left[- \sqrt{\frac{r}{r_g}} (r + 3 r_g)+3 r_g \tanh
	^{-1}\left( \sqrt{\frac{r}{r_g}}\right)+\sqrt{\frac{r_0}{r_g}}
	(r_0+3 r_g)-3 r_g \tanh
	^{-1}\left(\sqrt{\frac{r_0}{r_g}}\right)\right] \, , 
	\label{solminimal}
	\ee
	where $r_g=2 M$ is the Schwarzschild gravitational radius and $r_0$ is the initial condition, namely the point from which the particle starts its motion towards $r_g$. Near $r_g$, the solution (\ref{solminimal}) is approximated by:
	\be
	\tau(r) \approx \frac{r_g}{e} \log \left( \frac{r_0 - r_g}{r - r_g} \right) \, ,
	\ee
	which is in agreement with (\ref{finalGE3}). 
	
	For the case (\ref{incomplete}), the proper time to reach the surface $r = 2 M$ is finite and the spacetime turns out to be incomplete because the radial geodesic equation (\ref{finalGE}) is only defined for $r>2 M$, indeed, for $r=2M$ the function $S(r)$ in front of $\dot{r}^2$  in (\ref{finalGE}) is divergent. On the other hand, for (\ref{minimal}) the proper time to reach the surface $r = 2 M$ is infinity and, therefore, the spacetime is geodesically complete. 
	
	Looking at more general $S(r)$, it turns out that (\ref{minimal}), (\ref{red-blue2}), and (\ref{red-bluen}) 
	imply a logarithmic behavior of the proper time near the surface $r = 2 M$, while for (\ref{red-bluenm}) the behaviour is the inverse of a polynomial.

	\subsection{Geodesics for light rays or general massless particles}\label{luceSch}

	For massless particles, the correct action, which is invariant under reparametrization of the world line, $p^\prime= f(p)$, is: 
	\be
	S_{\gamma} = \int \mathcal{L}_{ \gamma } d \lambda = \int e(p)^{-1} \phi^2\hat{g}_{\mu\nu}  \frac{d x^\mu}{d p} \frac{d x^\nu}{d p} d p \, ,
	\label{Lm0}
	\ee
	where $e(p)$ is an auxiliary field that transforms as $e^\prime(p^\prime)^{-1} = e(p)^{-1} (d p^\prime/d p)$ in order to guarantee the invariance of the action. The action (\ref{Lm0}) is not only invariant under general coordinate transformations, but also under the Weyl conformal rescaling (\ref{Winv}) because of the presence of the dilaton field. 
	
	The variation with respect to $e$ gives:
	\be
	\frac{ \delta S_{\gamma}}{\delta e} = 0 \quad \Longrightarrow \quad  - \int d p  \, \frac{\delta e}{e^2} \phi^2 \, \hat{g}_{\mu\nu} \, \dot{x}^{\mu} \dot{x}^\nu = 0 \quad \Longrightarrow \quad   d \hat{s}^2 =  \hat{g}_{\mu\nu} dx^\mu d x^\nu = 0 \, ,
	\label{ds0}
	\ee
	consistently with the main property of massless particles or the equivalence principle.
	
	The variation with respect to $x^\mu$ gives the geodesic equation in the presence of the dilaton field, namely (in the gauge $e(p)={\rm const.}$)
	\be
	\frac{D^2(g = \phi^2 \hat{g}) x^\lambda}{d p^2} = 
	\frac{D^2(\hat{g}) x^\lambda}{d p^2} + 2 \frac{\partial_\mu \phi}{\phi} \frac{ d x^\mu}{d p} \frac{ d x^\lambda }{d p} 
	-  \frac{\partial^\lambda \phi }{\phi} \frac{ d x^\mu}{d p} \frac{ d x_\mu }{d p} 
	= 0 \, ,
	\label{geom0}
	\ee
	where $D^2(\hat{g})$ is the covariant derivative with respect to the metric $\hat{g}_{\mu\nu}$.
	
	However, when we contract equation (\ref{geom0}) with the velocity $d x_\lambda/d p$ and we use $d\hat{s}^2 =0$ obtained in (\ref{ds0}), we get the following on-shell condition, 
	\be
	\frac{d x_\lambda}{d p} \, \frac{D^2(\hat{g}) x^\lambda}{d p^2} + 2  \frac{d x_\lambda}{d p} \, \frac{\partial_\mu \phi}{\phi} \frac{ d x^\mu}{d p} \frac{ d x^\lambda }{d p} 
	-  \frac{d x_\lambda}{d p} \, \frac{\partial^\lambda \phi }{\phi} \frac{ d x^\mu}{d p} \frac{ d x_\mu }{d p} 
	= 0 \quad \Longrightarrow \quad \frac{d x_\lambda}{d p} \, \frac{D^2(\hat{g}) x^\lambda}{d p^2} = 0 \, .
	\ee
	
	Therefore, the $\frac{D^2(\hat{g}) x^\lambda}{d p^2}$ must be proportional to the velocity, namely 
	\be
	\frac{D^2(\hat{g}) x^\lambda}{d p^2} = f \, \frac{d x^\lambda}{d p}  \quad (f = {\rm const.}) 
	\label{constre}
	\ee
	because the velocity is null on the light cone. 
	Under a reparametrization of the world line $q = q(p)$, eq.~(\ref{constre}) becomes 
	\be
	\frac{d^2 x^{\lambda} }{dq^2} + \Gamma^\lambda_{\mu\nu}  \frac{d x^\mu }{d q}  \frac{d x^\nu }{d q} 
	=  \frac{d x^\lambda}{d p} \left( \frac{dp}{dq} \right)^2 \left(f \frac{dq}{dp} - \frac{d^2 q}{dp^2} \right) .
	\label{parame}
	\ee
	Choosing the dependence of $q$ on $p$ to make the right-hand side of (\ref{parame}) vanish, we end up with the geodesic equation in the affine parametrization. Hence, we can redefine $q \rightarrow \lambda$ and, finally, we get the affinely parametrized geodesic equation for photons in the metric $\hat{g}_{\mu\nu}$, 
	\be
	\frac{D^2(\hat{g}) x^\lambda}{d \lambda^2} = 0 \, .
	\label{affine}
	\ee
	
\begin{figure}
		\begin{center}
			\includegraphics[height=7cm]{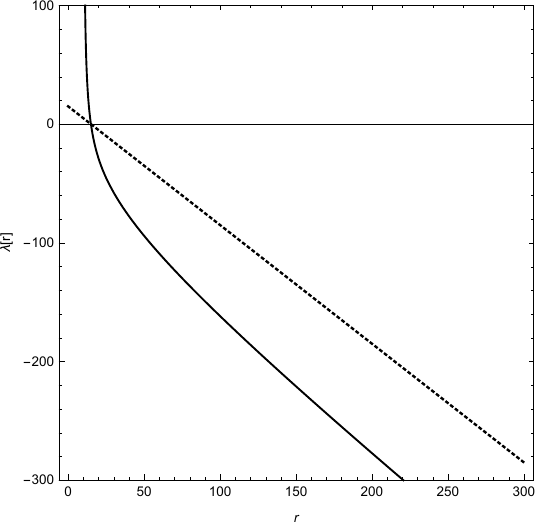}
			\hspace{2cm}
			\includegraphics[height=7cm]{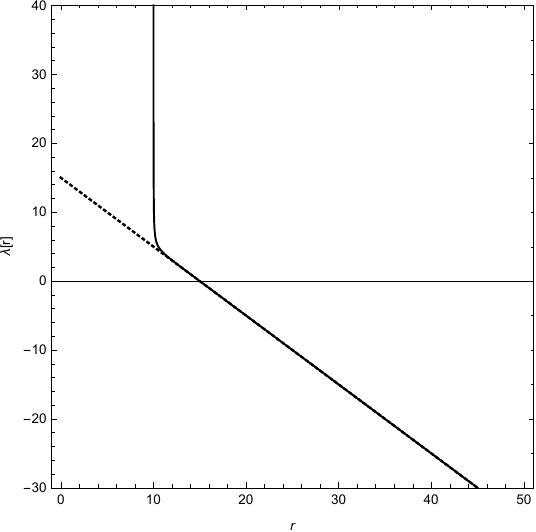}
		\end{center}
		\caption{{Left-plot: the affine parameter $\lambda(r)$ for the rescaling (\ref{minimal}), whose analytic expression is reported in (\ref{lambda}), for $e=1$, $r_0 = 3 M$, $M=5$. Right-plot: the affine parameter $\lambda(r)$ for the rescaling (\ref{red-blue2}) and $e=1$, $r_0 = 3 M$, $M=5$. The dotted lines in both plots correspond to the Schwarzschild case.}
		}
		\label{BLUESHIFT}
	\end{figure}

	We can now investigate the conservations laws based on the symmetries of the metric. Let us consider the following scalar,
	\be
	\hat{\alpha} = \hat{g}_{\mu\nu} v^\mu \frac{d x^\nu}{d \lambda} \, .
	\label{alpha}
	\ee
	where $v^\mu$ is a general vector. Taking the derivative of (\ref{alpha}) with respect to $\lambda$ and using the geodesic equation (\ref{affine}), we get
	
	\be
	\frac{d}{d \lambda} \hat{\alpha} = \frac{1}{2}  v^\mu \partial_\mu \hat{g}_{\rho \nu} \frac{d x^\rho}{d \lambda} 
	\frac{d x^\nu}{d \lambda} + \hat{g}_{\mu\nu} \partial_\rho v^\mu \frac{d x^\nu}{d \lambda} \frac{d x^\rho}{d \lambda} 
	= \frac{1}{2} [\mathcal{L}_v \hat{g} ]_{\rho \nu }  \frac{d x^\rho}{d \lambda} \frac{d x^\nu}{d \lambda} \, , 
	\ee
	where $[\mathcal{L}_v \hat{g}]$ is the Lie derivative of $\hat{g}_{\mu\nu}$ by a vector field $v^\mu$. 
	Thus, if $v^\mu$ is a Killing vector field, namely $[\mathcal{L}_v \hat{g}]=0$, $\hat{\alpha}$ is conserved:
	\be
	\frac{d}{d \lambda} \left[ \hat{g}_{\mu\nu} v^\mu \frac{d x^\nu}{d \lambda} \right] = 0 
	\, . 
	\label{CL}
	\ee
	The metric~(\ref{UBHmetric}) is time-independent and spherically symmetric (in particular, it is invariant under $t \rightarrow t + \delta t$ and $\varphi \rightarrow \varphi + \delta \varphi$). Therefore, we have the following Killing vectors associated with the above symmetries
	\be
	\xi^{\alpha} = (1, 0, 0, 0) \, , \quad \eta^{\alpha} = (0, 0, 0, 1) \, .
	\ee
	Since the metric is independent of the $t$- and $\varphi$-coordinates, we can construct the following conserved quantities
	\be
	&& e = - \xi \cdot u = - \xi^\alpha u^\beta \hat{g}_{\alpha \beta} = - \hat{g}_{ t \beta} u^\beta = - \hat{g}_{ t t } u^t = 
	S(r) \left(  1 - \frac{2 M}{r} \right) \frac{d t}{d \lambda} = S(r) \left(  1 - \frac{2 M}{r} \right) \dot{t} \, , 
	\label{Ldott} \\ 
	&& \ell = \eta \cdot u = \eta^\alpha u^\beta \hat{g}_{\alpha \beta} =  \hat{g}_{ \phi \beta} u^\beta =  \hat{g}_{ \phi \phi } u^\phi = 
	S(r) r^2 \sin^2 \theta \, \dot{\varphi} \, ,
	\label{Ldotphi}
	\ee 
	where the null vector 
	\be
	u^{\alpha} = \frac{d x^\alpha}{d \lambda} 
	\ee
	satisfies 
	\be
	u \cdot u = \hat{g}_{\alpha \beta}  \frac{d x^\alpha}{d \lambda}  \frac{d x^\beta}{d \lambda} = 0 \, .
	\label{NullV}
	\ee
	From~(\ref{NullV}), we get the following equation
	\be
	- \left(  1 - \frac{2 M}{r} \right) \dot{t}^2 + \frac{\dot{r}^2}{\left(  1 - \frac{2 M}{r} \right)} 
	+ r^2 \sin^2 \theta \dot{\varphi}^2 = 0 \, .
	\label{uu}
	\ee
	Note that the rescaling of the metric cancels out in the above equation \eqref{uu} for null geodesics, but $S(r)$ will appear again when the conserved quantities (\ref{Ldott}) and (\ref{Ldotphi}) are taken into account. Let us solve (\ref{Ldott}) for $\dot{t}$ and (\ref{Ldotphi}) for $\dot{\varphi}$ and, afterwards, replace the results in (\ref{uu}). The outcome is: 
	\be
	-  \frac{e^2}{S(r)^2 \left(1 - \frac{2 M}{r} \right)} 
	+ \frac{\dot{r}^2}{  1 - \frac{2 M}{r} } 
	+ \frac{\ell^2}{ S(r)^2 r^2 } = 0 \, .
	\label{uu2}
	\ee
	Let us focus on the radial geodesics (i.e. $\ell=0$), which will be sufficient to verify the geodesic completeness. Equation~(\ref{uu2}) simplifies, for $r > 2 M$, to 
	\be
	-  \frac{e^2}{S(r)^2 } 
	+\dot{r}^2 =0 \, \quad \Longrightarrow \quad S(r) | \dot{r} | = | e |\, .
	\label{gamma10}
	\ee
	For the sake of simplicity, from now on we assume $e>0$. 
	The above (\ref{gamma10}) first order differential equation can be easily integrated for a photon trajectory approaching $r_g= 2 M$, namely for $\dot{r} < 0$. For the rescaling (\ref{minimal}), one finds
	\be
	\lambda \, e = 2  \, r_g \log \left( \frac{r_0 - r_g}{r-r_g} \right)+r
	\left(\frac{r}{r-r_g}-2\right)-\frac{r_0 r_g}{r_0-r_g}+r_0 \, .
	\label{lambda}
	\ee
	Fig.~\ref{BLUESHIFT} shows $\lambda(r)$ for the rescaling factors in (\ref{minimal}) and (\ref{red-blue2}). 
	It turns out that photons cannot reach $r= 2 M$ for any finite value of the affine parameter $\lambda$.

	\begin{figure}
		\begin{center}
			\includegraphics[height=8cm]{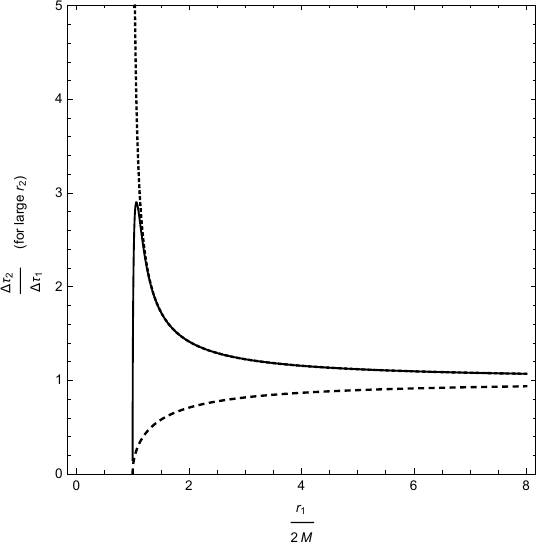}
		\end{center}
		\caption{{Plots of $\Delta \tau_2/\Delta \tau_1$ in the limit of large $r_2$: the dotted curve is for the Schwarzschild case ($\Delta \tau_2/\Delta \tau_1$ diverges when $r_1$ goes to $2M$), the dashed curve is for the rescaling (\ref{minimal}), and the solid curve is for the rescaling (\ref{red-blue2}).} 
		Notice that $\Delta \tau_2/\Delta \tau_1$ depends on the particular rescaling, but not on the mass of the black hole, as the plot for the rescaled radial coordinate shows. }
		\label{BLUESHIFT2}
	\end{figure}

	\section{Gravitational blueshift}
	Let us look at the time measured by two observers located, respectively, at the radial coordinates $r_1$ and $r_2  > r_1$. In the limit of very large $r_2$, we have 
	\be
	\frac{\Delta \tau_2}{\Delta \tau_1} = \sqrt{ \frac{ S(r_2) \left( 1 - \frac{r_g}{r_2} \right) }{ S(r_1) \left( 1 - \frac{r_g}{r_1} \right)}} \quad \rightarrow \quad \sqrt{ \frac{ 1 }{ S(r_1) \left( 1 - \frac{r_g}{r_1} \right)}}
	\label{BS}
	\, .
	\ee
	The main difference with respect to the black hole 
	case consists on a blueshift instead of a redshift as evident looking at the ratio (\ref{BS}). Indeed, for every rescaling except (\ref{incomplete}), (\ref{BS}) vanishes at $r = 2 M$, whereas it diverges in the black hole case. Notice that for (\ref{incomplete}) there is neither gravitational redshift nor gravitational blueshift because $\frac{\Delta \tau_2}{\Delta \tau_1} =1$.
	In Fig.~\ref{BLUESHIFT2} we compare the ratios $\Delta \tau_2/\Delta \tau_1$ evaluated for the rescalings (\ref{minimal}) and (\ref{red-blue2}), respectively.

	Moving now to the frequency issue, since the frequency is the inverse of proper time, namely $\omega = 1/\tau$, we get 
	\be 
	\omega_2 = 
	\omega_1 \, \sqrt{S(r_1) \left( 1 - \frac{r_g}{r_1} \right) } \, ,
	\ee
	where $\omega_1$ is the frequency measured in the proper time of the source emitting the light and $\omega_2$ the frequency measured by an observer at infinity.
	In the Schwarzschild case, when $r$ approaches $r_g$ the frequency seen by an observer at infinity goes to zero, while in the metric (\ref{UBHmetric}) the frequency goes to infinity, except for the rescaling (\ref{incomplete}). Indeed, the latter case is a particular choice of $n$ and $m$ in (\ref{red-bluenm}), namely $n=1$ and $m=1$, but let us consider the general case,
\be 
	\omega_2 = 
	\omega_1 \, \sqrt{\left(\frac{r_1^n}{r_1^n - r_g^n} \right)^m \left( \frac{r_1 - r_g}{r_1} \right) } \, ,
	\ee
	which implies $\omega_2 = \omega_1$ for the metric defined by (\ref{incomplete}) and $\omega_2/\omega_1 \rightarrow + \infty$ for $n=1$, $m>1$.

		Let us now consider the black hole emission of Hawking's particles that reach infinity with a particular (typically small) temperature $T_{\rm H}$ \cite{Hawking74}. When this temperature is traced back from infinity to the event horizon, it turns out to be Trans-Planckian at a Planck distance from $r_g$. Indeed, for the Schwarzschild metric we get:
		\be 
		\label{freqredshift}
		T_\infty = T_{\rm H} = 
		T_1 \, \sqrt{ \left( 1 - \frac{r_g}{r_1} \right) } \quad \Longrightarrow \quad 
		T_1 = \frac{T_{\rm H} }{
			\sqrt{ \left( 1 - \frac{r_g}{r_1} \right) }} \, \rightarrow + \infty \quad  {\rm for} \quad r_1 \rightarrow r_g 
		\, .
		\ee
		In the past, this argument made people question the validity of Hawking's original derivation that was based on a semi-classical approach \cite{Hawking74,tHooft:1984kcu,Unruh1,Jacobson,Brout,Helfer,Brandenberger:2011gk}.

		However, for conformalons, the presence of the rescaling can make the temperature (see next section) vanishes as we approach the surface $r = r_g$, 
		\be 
		\label{freqblueshift}
		T_\infty = T_{\rm H} = 
		T_1 \, \sqrt{S(r_1) \left( 1 - \frac{r_g}{r_1} \right) } \quad \Longrightarrow \quad 
		T_1 = \frac{T_{\rm H} }{
			\sqrt{S(r_1) \left( 1 - \frac{r_g}{r_1} \right) }}
		\, \rightarrow 0 \quad  {\rm for} \quad r_1 \rightarrow r_g 
		\, .
		\ee

		As we will show in the next section, 
		these objects have zero Hawking's temperature for the rescaling (\ref{incomplete}) or {\em formally} negative temperature for the other rescalings. We say {\em formally} because at this sage we do not want get in the interpretation of the result that we think to be extremely interesting in connection with the thermodynamics of systems at negative absolute temperature \cite{NegativeTemperature}.  
%
		 Therefore, according to (\ref{freqblueshift}) geodetically complete conformalons can easily solve the Trans-Planckian problem. 
		Once again, it is simply ``classical conformal geometry'' to solve an issue apparently traceable to quantum gravity effects.

		\begin{figure}
			\begin{center}
				\includegraphics[height=14cm]{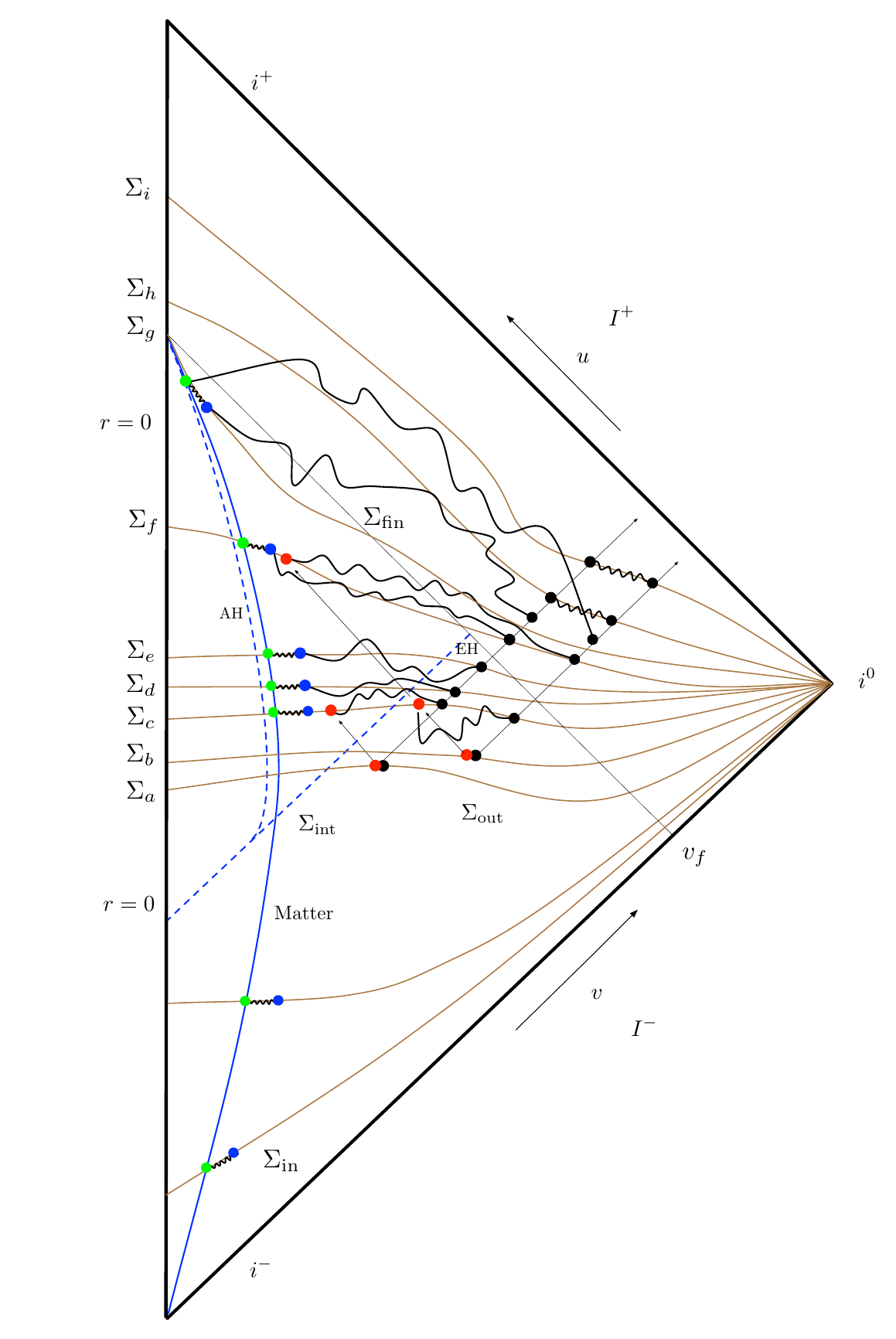}
			\end{center}
			\caption{The Penrose diagram for an evaporating conformalon --- This figure shows the transfer of entanglement from the collapsing matter (represented by the geodesic blue line for the pair of entangled particles shown 
				on the surface $\Sigma_{\rm in}$), which never forms the event horizon (dashed  straight blue line), to the Hawking particles  emitted from the conformalon. A Hawking pair is created on the Cauchy surface $\Sigma_a$ and evolves to the surface $\Sigma_c$ where we see two entangled pairs: the ``int'' (red) and ``out'' (black) Hawking particles on the right and two entangled black hole matter particles (green and blue). On $\Sigma_d$ one of the matter particles (the blue one) and the ``int'' particle (red) interact and generate a new particle  making a system of three entangled particles that evolve to $\Sigma_e$. On $\Sigma_f$ the remaining matter particles (green and blue) come very close to a new int-Hawking particle (red) created on $\Sigma_b$ and in $\Sigma_g$ they interact providing an entangled system of four particles: two very close the apparent horizon (AH), which never forms, and two far from the conformalon. Finally, assuming full annihilation 
				of the green and blue particles near the AH we end up with two ``out'' entangled particles on $\Sigma_h$ that evolve to $\Sigma_i$ towards infinity. In this black hole geometry, particles take an infinite amount of time to reach the AH and interact with the int-Hawking particles outside. Thus, the interaction can happen smoothly, with no problems caused by any singularity.}
			\label{PenroseConformalons}
		\end{figure}

		\section{Hawking temperature and evaporation time}
		\label{Hawking temperature}
		The Hawking temperature is computed as usual by studying a massless scalar field theory in the background geometry of what is in our case a conformalon. However, for consistency with (\ref{UBHmetric}) we also assume the scalar field to be conformally coupled. Indeed, the metric (\ref{UBHmetric}) is an exact solution of Einstein's conformal gravity only if the Weyl conformal symmetry is spontaneously, but not explicitly, broken. Therefore, the radial equation of motion for the scalar field near the event horizon reduces to the one in Minkowski spacetime like for the Schwarzschild metric, and Hawking's computation does not get any modification. Let us now expand on the latter statement. 
		
		In a conformally invariant theory a scalar field $\Phi$ transforms as the dilaton in (\ref{Winv}), namely 
		\be
		\Phi^\prime = \Omega^{-1} \Phi \, .
		\ee
		Therefore, the equation of motion for $\Phi$ and $\Phi^\prime$ are related as follows, 
		\be
		\left( \Box^\prime + \frac{1}{6} R^\prime \right) \Phi^\prime = \Omega^{-3} \left( \Box + \frac{1}{6} R \right) \Phi = 0 \, ,
		\label{INVeom} 
		\ee
		consistently with conformal invariance. 
		Hence the solution of the Klein-Gordon equation in a different conformal frame is simply the rescaling of the solution in the Schwarzschild spacetime. In particular in a conformalon's background the solution is:
		\be
		\hat{\Phi} = S^{-1/2}(r) \Phi \, , 
		\label{RescSol}
		\ee
		where $\Phi$ is the well known solution in the Schwarzschild background. Hence, given the solution $\Phi$ we get $\hat{\Phi}$
		by means of (\ref{RescSol}).

		For the sake of completeness, let us here remind the Hawking's derivation pointing out the similarities and differences whether the background is the Schwarzschild metric or the conformalon. 
		
		Following the Hawking's derivation, we consider a Schwarzschild black hole that forms from a collapsing null shell, at the instant 
		$v=v_0$. The spacetime region defined by $v < v_0$ is called the ``in'' region, while the one defined by $v > v_0$ will be referred to as the ``out'' region. Hence, we expand the (quantum) scalar field $\Phi$ in terms of wave modes and creation and annihilation operators. However, such expansion is not unique, one can expand in terms of the modes at $\mathcal{J}^-$, where the spacetime is Minkowski before the black hole's formation, or in terms of the modes at the future null infinity $\mathcal{J}^+$, in the black hole background. Namely,
		\be
		\Phi &= &\sum_k u_k^{\rm in} a_k^{\rm in} + u_k^{{\rm in}^*} a_k^{{\rm in} \dagger} \\
		&= &\sum_k v_k^{\rm out} b_k^{\rm out} + v_k^{{\rm out}^*} b_k^{{\rm out} \dagger}.
		\ee
		The ``in'' and ``out'' modes are related to each other through the so-called Bogoliubov transformations, similarly the ``in'' and ``out'' creation and annihilation operators, namely \cite{Fabbri} 
		\be  \label{waveBogol1}
		&& v_k^{\rm out} = \sum_{k'} \alpha_{k k'} u_{k'}^{\rm in}  + \beta_{k k'} u_{k'}^{{\rm in}^*}  ,\\
		&& b_k^{\rm out} = \sum_{k'} \alpha^*_{k k'} a_{k'}^{\rm in}  - \beta^*_{k k'} a_{k'}^{{\rm in}\dagger} \, . 
		\ee
		Therefore, the way to compute the number of Hawking particles produced in a (possible) black hole evaporation process is through the expectation value of the ``out''  number operator in an ``in'' vacuum state, i.e. 
		\be
		\langle 0_{\rm in} | N_k^{\rm out} |0_{\rm in} \rangle = \langle 0_{\rm in} | b_k^{\rm out \dagger} b_k^{\rm out}  |0_{\rm in} \rangle.
		\label{NumberOut}
		\ee
		In order to evaluate (\ref{NumberOut}), we need to express the ``out'' operators in terms of the ``in'' operators, which yields:
		\be
		\langle 0_{\rm in} | N_k^{\rm out} |0_{\rm in} \rangle & = & \langle 0_{\rm in} | \left( \sum_{k'} \alpha_{k k'} a_{k'}^{\rm in \dagger}  - \beta_{k k'} a_{k'}^{{\rm in}} \right) \left( \sum_{k''} \alpha^*_{k k''} a_{k''}^{\rm in}  - \beta^*_{k k''} a_{k''}^{{\rm in}\dagger}  \right) |0_{\rm in} \rangle  \\
		& = & \sum_{k'}|\beta_{k k'}|^2 .
		\ee
		The main challenge for the Schwarzschild case computation is to evaluate the function $\beta_{k k'}$.
		From (\ref{waveBogol1}), one can check that 
		\be
		\beta_{k k'} = \left( v_k^{\rm out} ,u_{k'}^{{\rm in}^*}\right) 
		=  i \int_{\Sigma} d \Sigma  \, \,  n^\mu \left(v_k ^{\rm out} \partial_{\mu} u_{k' }^{\rm in}  - u_{k'} ^{\rm in} \partial _ \mu v_k ^{\rm out} \right), \label{beta1}
		\ee
		where $n^\mu$ is a unit vector normal to the hypersurface $\Sigma$ 
		(here we refer the reader to our appendix for a sketch of the evaluation, and to \cite{Fabbri,Davies} for a full derivation of $\beta_{k k'}$).

		Now for the case of a conformalon, we need to prove  that in a theory with conformal symmetry, $\beta_{k k'}$ is invariant under a conformal transformation, namely 
	\be \hat \beta_{k k'}(r) = \beta_{k k'}(r) , 
	\label{BetaEquality} 
	\ee
	that would clearly imply that conformalons emit Hawking radiation as Schwarzschild black holes (but, as shown below, their temperature is different),
	\be | \hat \beta_k|^2 = |\beta_k|^2 ,
	\ee
	where $k$, as already established, is the momentum of the created Hawking's particles at $\mathcal{I}^+$. 
	The Bogoliubov coefficient $\hat \beta_{k k'}$ for the rescaled metric, i.e. in the  conformalon background, is:
	\be 
	\label{beta} 
	\hat \beta_{k k'} = - (\hat u_k ^{\rm out},  \hat u_{k' }^{\rm in *}) 
	= i \int_{\Sigma} d \hat \Sigma \, \,  \hat g^{\mu \nu}\hat n_\nu \left(\hat u_k ^{\rm out} \partial_{\mu} \hat u_{k' }^{\rm in *}  -  \hat u_{k'} ^{\rm in *} \partial _ \mu \hat u_k ^{\rm out} \right) \, . 
	\ee
	Therefore, we need to know how each quantity in the above integral rescales under a conformal transformation. 
	Let us consider first the two terms inside the brackets, i.e. 
	\be 
	\hat u_k ^{\rm out} \partial_{\mu} \hat u_{k' }^{\rm in *}  -  \hat u_{k'} ^{\rm in *} \partial _ \mu \hat u_k ^{\rm out} 
	& = & \Omega^{-1} u_k ^{\rm out} \, \partial_{\mu} \left(\Omega^{-1} u_{k' }^{\rm in *}  \right)  -  \Omega^{-1} u_{k'} ^{\rm in *} \, \partial _ \mu \left( \Omega^{-1} u_k ^{\rm out} \right)  \nonumber \\
	&& \hspace{-2cm} 
	=   \Omega^{-1} u_k ^{\rm out} \, \partial_{\mu} \left( \Omega^{-1} \right) u_{k' }^{\rm in *}    
	+  \cancel{\Omega^{-1} u_k ^{\rm out}\, \Omega^{-1} \, \partial_{\mu}  \left( u_{k' }^{\rm in *}  \right) }
	-  \Omega^{-1} u_{k'} ^{\rm in *} \, \partial _ \mu \left( \Omega^{-1} \right) u_k ^{\rm out}  
	-  \cancel{\Omega^{-1} u_{k'} ^{\rm in *} \Omega^{-1} \, \partial _ \mu \left(  u_k ^{\rm out} \right)}
	\nonumber \\
	&& \hspace{-2cm}
	=   \Omega^{-2} \left( u_k ^{\rm out} \partial_{\mu} u_{k' }^{\rm in *}   -  u_{k'} ^{\rm in *} \partial _ \mu u_k ^{\rm out} \right),
	\label{braket}
	\ee
	which transforms as a conformal field of weight minus two. 
	
	On the other hand, the $3-$dimensional hypersurface integral measure scales as the square root of the determinant of the induced metric, namely 
	\be d \hat \Sigma = \sqrt{\Omega^6} d\Sigma = \Omega^3 d\Sigma \, .
	\ee
	Finally, we can figure out the scaling of the vector $\hat{n}^\mu$ according to its normalization, namely
	\be 
	1 = \hat n_\mu \hat n ^\mu =  \hat g^{\mu \nu }  \hat n _\mu  \hat n _\nu = \Omega^{-2}g^{\mu \nu }  \hat n _\mu  \hat n _\nu 
	= \Omega^{-2} \, g^{\mu \nu }\,  \Omega^{h_n}  \,  n _\mu \, \Omega^{h_n} \, \hat n _\nu  
	= \Omega^{-2} \,  \Omega^{h_n}  \,  \Omega^{h_n} \, \underbrace{g^{\mu \nu } n _\mu \, \hat n _\nu }_{= 1} 
	\quad \Longrightarrow \quad h_n = 1 .
	\label{measure}  
	\ee
	Therefore, $\hat n^\mu $ is correctly normalized if: 
	\be \hat n ^\mu = \Omega \, \,n^\mu.
	\label{confNorm}
	\ee
	Replacing now (\ref{braket}), (\ref{measure}), and (\ref{confNorm}) in (\ref{beta1}), 
	\be  
	\label{beta2} 
	\hat \beta_{k k'}&&	=  i \int_{\Sigma} d \hat \Sigma \, \,  \hat g^{\mu \nu}\hat n_\nu \left(\hat u_k ^{\rm out} \partial_{\mu} \hat u_{k' }^{\rm in *}  -  \hat u_{k'} ^{\rm in *} \partial _ \mu \hat u_k ^{\rm out} \right) \nonumber \\ 
	&& =  i \int_{\Sigma} \Omega^3  \, d \Sigma \, \,  \Omega^{-2} g^{\mu \nu} \, \Omega  \, n_\nu  \, \Omega^{-2} \left(\hat u_k ^{\rm out} \partial_{\mu} \hat u_{k' }^{\rm in *}  -  \hat u_{k'} ^{\rm in *} \partial _ \mu \hat u_k ^{\rm out} \right)  \\
	&&  =  i \int_{\Sigma}   d \Sigma \, \,   g^{\mu \nu }n_\nu  \left(u_k ^{\rm out} \partial_{\mu} u_{k' }^{\rm in *}  - u_{k'} ^{\rm in *} \partial _ \mu u_k ^{\rm out} \right) 
	=  \beta_{k k'}. 
	\ee  
	We conclude that the 
	beta Bogoliubov coefficient is conformally invariant. 
	
	The Hawking temperature for conformalons can be inferred by calculating the surface gravity of a conformalon at $r = 2 M$. 
		We start with the rescaled Schwarzschild metric in the advanced Eddington-Finklestein coordinates, namely 
		\be
		(v=t+r_*, r , \theta, \phi) \, , \quad \mbox{where} \quad  r_* = r + 2 M \ln \frac{|r- 2 M|}{2M} \, , 
		\label{Tortoise}
		\ee
		in which the metric reads:
		\be 
		d \hat{s}^2= S(r) \left[ - \left( 1-\frac{2M}{r} \right) dv^2 +dv dr + r^2 d \Omega^2 \right]  \, .
		\ee
		The metric has the Killing vector $\partial_v$ that we can represent as $k^\alpha = (1,0,0,0)$, and the covariant vector reads:
		\be 
		k_\alpha = g_{\alpha \beta} k^\beta = \bigg( - S(r) \left( 1-\frac{2M}{r} \right), \, S(r), \, 0, \, 0\bigg) \, .
		\ee
		The surface gravity $\kappa$ is then defined by the following relation, 
		\be 
		- k^\alpha \nabla _ \beta k _\alpha = \kappa \,  k_\beta \, . 
		\ee
		Evaluating the equation above for the index $\beta=v$ we get: 
		\be
		- k^\alpha  \Gamma_{\alpha v} ^\gamma k_{\gamma} = \kappa \, k_v \, . 
		\ee
		Since, the only relevant non-zero component of the Christoffel symbols is: 
		\be \Gamma_{vv}^v = \frac{1}{2S} \left[ \partial_r S(r) \left( 1- \frac{2M}{r} \right) + \frac{2M S(r)}{r^2} \right] \, .
		\ee
	%
Finally, the surface gravity for a general rescaling reads:
		\be \label{Sman345}
		\kappa = \frac{1}{2S(r)}   \left[  \partial_r S(r) 
		\left( 1 - \frac{2M}{r} \right)  + \frac{2M S(r) }{r^2} \right] \Bigg|_{r=2M} 
		\, .
		\ee
Another equivalent but shorter derivation of the latter statement consists on rewriting the surface gravity at the event horizon in terms of the components of the metric, namely $\hat{g}_{00}$ and $\hat{g}_{11}$, 
		\be
		\kappa& = & - \lim_{r \rightarrow 2M} \frac{1}{2} \frac{\partial_r \hat{g}_{00}}{\sqrt{ - \hat{g}_{00} \hat{g}_{11}}} 
		=  \lim_{r \rightarrow 2M} \frac{1}{2} \frac{\partial_r \left[ S(r) \left(1- \frac{2 M}{r} \right) \right] }{\sqrt{ S^2(r)}}
		=  
		\lim_{r \rightarrow 2M} \frac{1}{2} \frac{ \left( \partial_r S(r) \right) \left(1- \frac{2 M}{r} \right)  + S(r) \frac{2 M}{r^2} }{\sqrt{ S^2(r)}} \nonumber \\
		& = & 
\lim_{r \rightarrow 2M} \left[ \frac{1}{2} \frac{ \left( \partial_r S(r) \right) }{ \sqrt{ S^2(r)} }  \left(1- \frac{2 M}{r} \right)
		+ \frac{1}{2} \frac{  S(r) }{ \sqrt{ S^2(r)} }  \frac{2 M}{r^2} 
		\right] \, .
		\label{Sman}
		\ee
		In general, the surface gravity is conformally invariant if the rescaling $S(r)$ is an analytic function at the horizon. This is evident from (\ref{Sman345}) and the last equality in (\ref{Sman}). 
However, such a conclusion does not apply here because $S(r)$ is singular at the event horizon.  
In particular, for the conformal factor (\ref{red-bluen}) the surface gravity is negative and exactly the opposite of the Hawking temperature for all values of the integer $n$,
\be
\kappa= -  \frac{1}{4M} . 
\ee
For the rescaling (\ref{red-bluenm}) with $n=1$, i.e. 
\be
S(r) =  \left( \frac{r}{r - 2 M} \right)^m \, , 
\ee
the surface gravity is:
\be
\kappa= \frac{1 - m}{4M} . 
\ee
and it is positive only for $m<1$, while it is zero for $m=1$ which coincides with the rescaling (\ref{incomplete}). Notice that the temperature of the non-geodetically complete conformalons is positive but smaller than the Hawking's temperature for $m<1$. On the other hand, for $m>1$, the temperature is, at least formally, negative. However, while a negative temperature looks exotic, it has been known for a long time that thermodynamics at negative temperature is well defined \cite{NegativeTemperature} for those systems that exhibit an upper limit to energy. For such systems, the entropy is zero, and the temperature vanishes, in two cases: when all particles takes the minimum value of the energy or when all of them take the maximum value. In such systems, if one of the particles passes to a lower energy value the entropy increases and the temperature becomes smaller than zero.

Although a more in-depth conceptual and mathematical analysis is needed and a matter of a future work, we report here some results concerning the evaporation of conformalons exhibiting negative temperature.

    	The Hawking temperature for the conformalon, which is proportional to the surface gravity, is given by:
		\be
		T_{\rm H} = \frac{\kappa}{2 \pi} = \frac{1-m}{8 \pi M} \, ,
		\ee 
and as mention above negative for $m>1$.

		Let us now focus on the the area of the event horizon, 
		\be
		A_{\rm H} = \lim_{r \rightarrow 2 M} 4 \pi S(r) r^2, 
		\ee
		which is divergent for $r = 2 M$. However, according to the geodesic motion, explicitly studied in the previous section, nothing can reach the event horizon. Hence, the above divergence in the area is unattainable and the Hawking particles will be created near the horizon, but not exactly at the horizon in conformalons. This is not different from the original Hawking's derivation where the particle are created near the horizon but not at the horizon. Indeed, what we need in the original Hawking's computation is simply that the effective potential is nearly zero (see the appendix).

		Now, according to the Boltzmann law, we can evaluate the evaporation time integrating the following equation,
		\be
		- \frac{ d M}{d t} & = & \sigma A_{\rm H} \, T_{\rm H}^4 = \sigma \, S[2 M (1 + \delta)] 4 \pi [2 M (1+ \delta)]^2 \left( \frac{1}{8 \pi M} \right)^4 \, ,
		\label{EvapEq}
		\ee
		where the infinitesimal dimensionless parameter $\delta$ has been introduced to regularize the event horizon's area
		because most of the Hawking particles are created near -- but not at -- the horizon. Indeed, we stress once again that nothing can reach the event horizon in a finite amount of proper time (or for finite values of the affine parameter) as proved in the previous section. 
		It deserves to be notice that the evaporation time is unaffected by the negative value of the temperature. 
		
		As a particular example, eq.~(\ref{EvapEq}) for the rescaling (\ref{minimal}) takes the form:
		\be
		- \frac{ d M}{d t} = 
		\frac{\sigma (1+\delta )^6  }{256 \pi ^3 \delta^4} \frac{1}{ M^2}, 
		\ee
		and the solution is:
		\be
		M(t) = (M_0^3 - k \, t)^{1/3} \, , \qquad k = \frac{256 \pi ^3 \delta ^4}{3 (1+\delta)^6 \sigma }  
		\equiv    \frac{ \delta ^4}{ (1+\delta )^6  } \,  k_{\rm Sch.} \approx  \delta ^4 \,  k_{\rm Sch.}
		\, , \qquad t = \frac{1}{k} \left( M_0^3 - M^3 \right) 
		\,  .
		\ee 
		The evaporation time is finite like for the Schwarzschild case, but much longer because $\delta \ll 1$, namely particles created very near the horizon will take very long time to reach the observer at infinity. In order to make more accurate the latter statement, one should express the distance $\delta$ in terms of the time $t$ for the observer at infinity using the geodesics for massless particles. Technically we should solve (\ref{EvapEq}) with $\delta$ a function of time, namely we have to replace (\ref{EvapEq}) with:
		\be
		- \frac{ d M}{d t} = 
		\frac{\sigma (1+\delta(t) )^6  }{256 \pi ^3 \delta(t)^4} \frac{1}{ M^2} .
		\label{EvapEqt}
		\ee
		
		A similar result will turn out for the other rescalings too. 
		So far we have proved that the evaporation time of the conformalons is finite, but longer, than for the Schwarzschild black hole. The latter result could have implications also for the dark matter content of the Universe. 
		
		The thermal entropy defined through the following expression, 
		\be
		S_{\rm H} = \int \frac{ \delta Q}{T} = \int \frac{dM}{T}, 
		\ee
		is now positive when we extract energy from the system namely for $\delta Q < 0$ and $T<0$ (see the above short discussion). 
		\be
		S_{\rm H}  = - \int \frac{ \delta Q}{T} = - \int \frac{dM}{T}  
		= - \int d M \frac{8 \pi  M}{1 - m}
		=
		\frac{4 \pi  M^2}{m-1} \, ,
		\label{entropy}
		\ee
		which is positive for $m>1$. Notice that for large $m$ the entropy is smallar then $A_{\rm EH}/4$ consistently with the Bekenstein bound.
		When (\ref{entropy}) is expressed in terms of the event horizon's area $A_{\rm EH}$, using  the form factor (\ref{minimal}) ($m=2$ in  (\ref{entropy})), we get:
		\be
		S_{\rm H} = \frac{A_{\rm EH}}{4} \frac{\delta^2}{(1+\delta)^2} \approx  \frac{A_{\rm EH}}{4} \, \delta^2 \, ,
		\label{entropyA}
		\ee
		which looks much smaller then $A_{\rm EH}/4$ for small $\delta$. However, the event horizon's area is implicitly a function of $\delta$ that diverges for $\delta \rightarrow 0$, i.e.  
		\be
		\lim_{\delta \rightarrow 0} A_{\rm EH} = + \infty ,
		\ee
		consistently with the finiteness of the entropy (\ref{entropy}), which is expressed in terms of the mass.

		\section{Annihilation of the black hole matter particles}
\begin{figure}
			\begin{center}
				\includegraphics[width=0.6\linewidth]{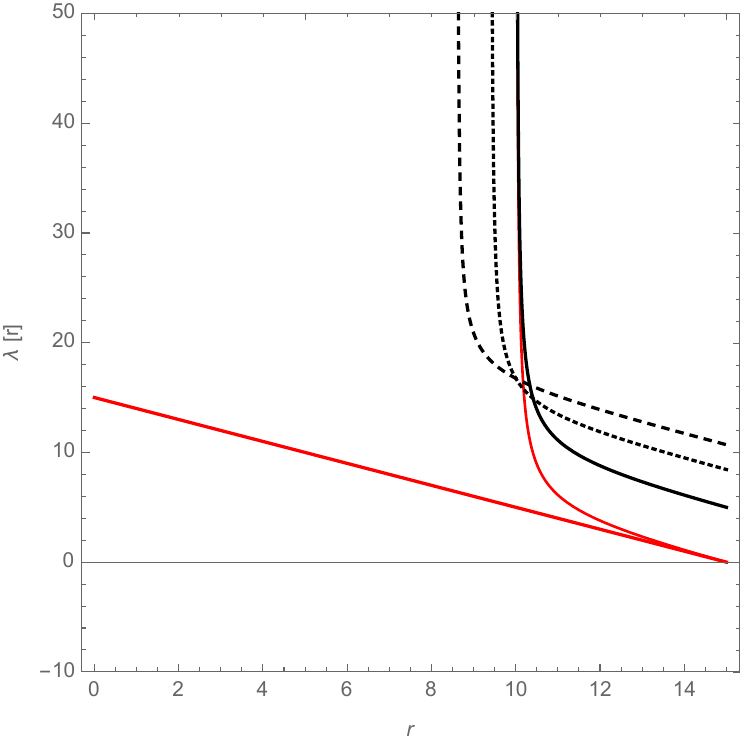}
			\end{center}
			\caption{Here we show the affine parameter $\lambda(r)$ for several geodesics. We have chosen the constant $e=1$ because it can always be absorbed in a finite rescaling of the affine parameter $\lambda$. 
    The initial mass of the black hole is $M_1=5$ and the event horizon is located in $r_g = 2 M_1 = 10$ while the initial value for the affine parameter is set to be zero in $r_0 = 3 M_1 = 15$ for the collapsing shell, namely 
    $\lambda(r_0 = 3 M_1 = 15) =0$. The two red geodesics represent $\lambda(r)$ of the collapsing matter for the Schwarzschild case and for the rescaling (\ref{red-bluen}) in which $n=8$. 
    For the Hawking-int particles created later we have chosen the integration constants such that 
    $\lambda(r_0 = 3 M_1) = M_1 = 5$,
    $\lambda(r_0 = 3 M_2) = 2 M_2 = 2 \times 4.7$, 
    $\lambda(r_0 = 3 M_3) = 3 M_3 = 3 \times 4.3$
    respectively. Notice that the solid black line represents a particle with negative energy that can never hit the collapsing shell because it is created when the mass of the black hole is exactly the initial one. However, other int-particles created later, when the mass of the black is smaller, can hit such particle creating a new particle having even more negative energy, but now able to hit the collapsing matter.
				\label{Hit}
			}
		\end{figure}

		In the spacetimes introduced in this paper, namely the conformalons, the Hawking process takes place outside the event horizon where int-particles (of negative energy) and collapsing (entangled in Fig.~\ref{PenroseConformalons}) matter interact near the surface $r = 2 M$, that is never achieved, till the complete exhaustion of the collapsing fuel.
		The whole process is explained in the caption of Fig.~\ref{PenroseConformalons}. 
		The latter statement is proved looking at the geodesics of light for (as an example) the rescaling factor (\ref{red-bluen}) and taking $n=8$, namely
		\be
		\lambda(r) \, e &  = &
		\frac{1}{128} \left(\frac{16 r {r_g}^8}{r^8- {r_g}^8}-9 \sqrt{2} {r_g} \log
		\left(r^2-\sqrt{2} r {r_g}+{r_g}^2\right)+9 \sqrt{2} {r_g} \log \left(r^2+\sqrt{2}
		r {r_g}+{r_g}^2\right)-18 {r_g} \log ({r_g}-r)
		\right. 
		\nonumber \\
		&&
		\left. 
		+18 {r_g} \log
		(r+{r_g})+18 \sqrt{2} {r_g} \tan ^{-1}\left(\frac{\sqrt{2} r}{{r_g}}+1\right)+36
		{r_g} \tan ^{-1}\left(\frac{r}{{r_g}}\right)+18 \sqrt{2} {r_g} \tan
		^{-1}\left(\frac{\sqrt{2} r-{r_g}}{{r_g}}\right)-128 r \right)
		\nonumber \\
		&&
		- ( r \rightarrow r_0)  
		\, ,
		\ee
		which should be compared with the geodesic in the Schwarzschild metric (namely in the Schwarzschild conformal frame),
		\be
		\lambda(r) \, e = - r + r_0 \, . 
		\ee
		In Fig.~\ref{Hit} we plot several geodesics for $r_0 = 3M$, but representing Hawking int-particles created later. An initial pure state evolves into a final pure state according to the analysis explained in~\cite{Akil:2018iny}.

		\section{Conclusions}
		
		The solutions of a general class of conformally invariant theories could be proposed as an alternative to black holes. The spacetime metric is obtained by means of a conformal rescaling of the Schwarzschild metric, which is singular at the event horizon. The latter property rules out the black hole interior from our Universe, as proved looking at the dynamics of probe-particles in the resulting geometry. Indeed, neither massive nor massless particles can reach the surface of the Schwarzschild event horizon. 

		Conformalons can somehow remind gravitational instantons in Euclidean general relativity in which the Wick rotation excludes the black hole interior from the spacetime. 
		Therefore, conformalons are vacuum solutions that spontaneously break the Weyl symmetry, but preserve the general coordinate invariance. Indeed, all the observables will be invariant under diffeomorphisms.
		There are obviously multiple choices of this kind of vacuum solutions, but all of them with the same properties, that is: unreachable horizon, same causal structure, singularity-free because there is no black hole's interior.
	However, a very new feature is about the negative value of the surface gravity, and, accordingly, of the Hawking's temperature. This is a very interesting feature that we have here only marginally touched but we hope to investigate more in the near future in connection with the thermodynamics at negative temperature \cite{NegativeTemperature}. 
		
		A very peculiar property of the conformalons is the gravitational blueshift instead of gravitational redshift. This is a remarkable difference with respect to the Schwarzschild spacetime. From the theoretical side, it provides a simple solution to the Trans-Planckian problem because they have {\em formally} negative temperature, namely they are stable at quantum level.

	Another more observational motivation for conformalons in place of black holes is related to a recent paper \cite{Faraoni} where it is shown under quite general assumptions that black holes turns out to be singular at the event horizon in a dynamical Universe. At the moment, conformalons provide a possible solution of this issue because the event horizon becomes an asymptotic Kretschmann-flat (\ref{K}) boundary region.

		From the phenomenological point of view, conformalon metrics could be tested with observational data of astrophysical black holes. For example, the analysis of the reflection features commonly observed in the X-ray spectrum of accreting black holes~\cite{Bambi:2016sac,Tripathi:2020yts,Bambi:2020jpe} can presumably rule out some choices of the rescaling $S(r)$, as some of them may not be able to predict the very broadened iron lines observed in some sources. However, this would require a detailed analysis which is beyond the scope of the present paper and is postponed for the future. We cannot easily measure the redshift/blueshift at a specific radial coordinate. From observations, we can only measure the spectrum of the radiation emitted from an extended region of the accretion disk and the gravitational redshift/blueshift is mixed together with the Doppler redshift/blueshift due to the motion of the material emitting the radiation and the light bending due to the strong gravitational field around the compact object~\cite{Bambi:2017khi}.

		\vspace{0.5cm}

		{\bf Acknowledgments --}
		{We thank the anonymous referee for noticing a mistake in our calculation of the Hawking temperature. L.M. would like also to thank Mariano Cadoni for useful discussions and suggestions.}

		\section*{Appendix: Particle Creation trough the formation of a Black Hole}
		We here review the standard derivation of the Hawking evaporation phenomenon within all the details. Note that we will heavily rely on the book by Fabbri and Salas \cite{Fabbri}.
		Let us consider a massless dust shell that is collapsing to form a black hole of mass $M$. Before the collapse, we assume the spacetime to be Minkowski, while after the collapse the spacetime is described by the Schwarzschild metric. Therefore, any quantum instability will be due to the matching boundary conditions. 

		We denote the Minkowski's patch of the spacetime as the ``in" region, and the Schwarzschild patch as the ``out" region. Hence, the ``in" region is described by the metric:
		\be 
		ds^2 _{\rm in} = -dt_{\rm in}^2 + dr_{\rm in}^2 +r_{\rm in}^2 d \Omega^2 
		= - du_{\rm in} d v + r_{\rm in}^2 d \Omega^{(2)} 
		\, , 
		\label{Mink}
		\ee 
		where we introduced the advanced and retarded coordinates for the Minkowski spacetime, namely
		\be
		u_{\rm in} = t_{\rm in} - r_{\rm in} \quad  {\rm and} \quad  v = t_{\rm in} + r_{\rm in}.
		\label{uvM}
		\ee
		On the other hand the ``out" region is described by:
		\be 
		ds^2 _{\rm out} = 
		- \left( 1-\frac{2M}{r_{\rm out}} \right) d u_{\rm out} dv + r_{\rm out}^2 d \Omega^2  \, ,
		\label{Schout}
		\ee 
		where we have introduced the advanced and retarded coordinates for the Schwarzschild spacetime, i.e. 
		\be 
		u_{\rm out} = t_{\rm out} - r_{*\rm out}  \quad  {\rm and} \quad v = t_{\rm out} + r_{* \rm out},
		\label{uvS}
		\ee 
		and the the tortoise coordinate $r_*$ has been defined in (\ref{Tortoise}).

		Now we consider the equation of motion of a scalar field on a general background (later we will specify to the Schwarzschild's metric),
		\be
		\left( \Box + \frac{1}{6} R \right)\Phi(x)=  \frac{1}{  \sqrt{-  g}}  \partial_\mu \big[ \sqrt{- g} g^{\mu \nu} \partial_\nu  \Phi(x) \big]   +  \frac{1}{6} R \,  \Phi(x)   = 0 \, .
		\label{INVeom2} 
		\ee
		The quantum scalar field can be expanded in terms of wave modes, and creation, annihilation operators. There is no unique basis for such expansion, but there are choices relevant for the whole process of collapse and evaporation, namely: the modes at  $\mathcal J^-$, the past null-infinity, and those at  $\mathcal J^+$, the future null-infinity.  Hence, the derivation of the Hawking radiation boils down to compare those two sets of modes, i.e. 
		\be
		\Phi = \sum_k u_k^{\rm in} a_k^{\rm in} + u_k^{{\rm in}^*} a_k^{{\rm in} \dagger} 
		= \sum_k v_k^{\rm out} b_k^{\rm out} + v_k^{{\rm out}^*} b_k^{{\rm out} \dagger},
		\ee
		where by ``in" and ``out" we mean the regions before and after the black hole formation, respectively. $u_k^{\rm in}$ and the $v_k^{\rm out}$ are the wave modes with momentum $k$, whereas $a_k^{\rm in}$ and $b_k^{\rm out}$ are the annihilation operators in the ``in" and ``out" regions respectively. 
		As already mention on the main text, the ``in" and ``out" modes are related to each other through the so-called Bogoliubov transformations. The same holds for the ``in" and ``out " creation and annihilation operators \cite{Fabbri}, 
		\be  \label{waveBogol}
		v_k^{\rm out} &=& \sum_{k'} \alpha_{k k'} u_k^{\rm in}  + \beta_{k k'} u_k^{{\rm in}^*}  ,\\
		b_k^{\rm out} &= & \sum_{k'} \alpha^*_{k k'} a_k^{\rm in}  - \beta^*_{k k'} a_k^{{\rm in}\dagger} \, 
		\ee
		where $u_k^{\rm in}$ is a solution to the Klein-Gordon equation in the Minkowski background, whereas $v_k ^{\rm out}$ is a solution of the Klein-Gordon equation in the Schwarzschild background. %
		Both the solutions of the Klein-Gordon will be shortly derived. 

		Let us now compute the number of Hawking particles produced by a Schwarzschild black hole, namely the expectation value of the ``out" number operator in the ``in" vacuum state. That is,
		\be
		\langle 0_{\rm in} | N_k^{\rm out} |0_{\rm in} \rangle = \langle 0_{\rm in} | b_k^{\rm out \dagger} b_k^{\rm out}  |0_{\rm in} \rangle.
		\label{enne}
		\ee
		In order to evaluate the above expectation value, we need to express the ``out" operators in terms of the ``in" operators, namely:
		\be
		b_k^{\rm out \dagger}  =  \sum_{k'} \alpha_{k k'} a_k^{\rm in \dagger}  - \beta_{k k'} a_k^{{\rm in}} \, , 
		\qquad 
		b_k^{\rm out}   =   \sum_{k''} \alpha^*_{k k''} a_k^{\rm in}  - \beta^*_{k k''} a_k^{{\rm in}\dagger} \, .
		\label{bdb}
		\ee
		Hence (\ref{enne}) reads:
		\be
		\langle 0_{\rm in} | N_k^{\rm out} |0_{\rm in} \rangle &=&  \langle 0_{\rm in} | \left( \sum_{k'} \alpha_{k k'} a_k^{\rm in \dagger}  - \beta_{k k'} a_k^{{\rm in}} \right) \left( \sum_{k''} \alpha^*_{k k''} a_k^{\rm in}  - \beta^*_{k k''} a_k^{{\rm in}\dagger}  \right) |0_{\rm in} \rangle  
		=  \sum_{k'}|\beta_{k k'}|^2 . \label{particNumb}
		\ee
		The main challenge, in the Schwarzschild background, is to evaluate the function $\beta_{k k'}$, which  
		it is related to the solutions of the EoM by means of (\ref{waveBogol}). The result is:
		\be
		\beta_{k k'} =  \left( v_k^{\rm out} ,u_{k'}^{{\rm in}^*}\right) \label{betaInner} 
		= i \int_{\Sigma} d \Sigma  \, \,  n^\mu \left(u_k ^{\rm out} \partial_{\mu} u_{k' }^{\rm in}  - u_{k'} ^{\rm in} \partial _ \mu u_k ^{\rm out} \right).
		\ee

		The last effort consists on solving the Klein-Gordon equation (\ref{INVeom2}) in the ``int'' and ``out'' regions. 
		
		For the sake of simplicity, we remove the labels ``in" (``out") for both the radial and the time coordinate in the Minkowski (Schwarzschild) spacetime. 
		
		Let us now solve the EoM for a scalar field in the Minkowski and in the Schwarzschild spacetime. Afterwards, we will express the two solutions in the same coordinates in order to evaluate (\ref{betaInner}). 

		{In the \underline{``int'' region, we have the Minkowski's metric} and the equation (\ref{INVeom2}) reads,
			\be 
			- \partial^2_t u_k^{\rm in}(x)  + \frac{1}{r^2 } \partial_r \left( r^2 \partial_r  \, u_k^{\rm in}(x)   \right)  + \frac{1}{r^2}  \frac{1}{\sin \theta} \partial_{\theta} \left( \sin \theta \, \partial_\theta u_k^{\rm in}(x)   \right)  + \frac{1}{r^2 \sin^2 \theta} \partial^2_{\varphi}  u_k^{\rm in} (x) 
			+ \frac{1}{6} R \, u_k^{\rm in} (x) = 0 ,
			\ee 
			where we noted $u_k^{\rm in} (x) \equiv u_k^{\rm in} (r,t,\theta,\varphi) $, and
			where the angular momentum operator reads: 
			\be 
			L^2  = \frac{1}{\sin \theta} \frac{\partial}{\partial \theta} \left( \sin \theta \frac{\partial}{\partial_\theta } \right)  + \frac{1}{\sin ^2 \theta} \frac{ \partial^2}{\partial \varphi ^2}  \, .
			\label{angular}
			\ee 
			The eigenfunctions of the operator (\ref{angular}) are the spherical harmonics $ Y_l(\theta, \varphi)$ with eigenvalues $-l(l+1)$. 
			Therefore, making use of the following general replacement, 
			\be
			u_k^{\rm in}(t,r,\theta,\varphi)= \sum_{l , m} \frac{ u^{\rm in}_{k,l}(t,r)}{r} Y_l (\theta,\varphi) \equiv
			\sum_{l , m}  u^{\rm in}_{k,l}(t,r)Y_l (\theta,\varphi)
			\, , 
			\ee
			we get the following reduced equation for $ u^{\rm in}_{k,l}(t,r)$, 
			\be 
			\left(- \frac{\partial^2}{\partial t ^2} +  \frac{\partial^2}{\partial r ^2} - \frac{l(l+1)}{r^2} \right)   u^{\rm in}_{k,l}(t,r) +\frac{1}{6} R \,  u^{\rm in}_{k,l}(t,r)= 0 .
			\label{frt}
			\ee 
			Since the Ricci scalar is zero in the Minkowski spacetime, (\ref{frt}) further simplifies to:
			\be 
			\left(- \frac{\partial^2}{\partial t ^2} +  \frac{\partial^2}{\partial r ^2} - \frac{l(l+1)}{r^2} \right) u^{\rm in}_{k,l}(t,r) = 0 \, ,
			\ee 
			which in the s-wave approximation ($l =0$) boils down to:
			\be \left(- \frac{\partial^2}{\partial t ^2} + \frac{\partial^2}{\partial r ^2}  \right)  u^{\rm in}_{k,0}(t,r) = 0 \, .
			\label{KGMin}
			\ee 
			The solution of (\ref{KGMin}) for the mode $k$ can be factorized as follows, 
			\be
			u^{\rm in}_{k,0}(t,r)  = {\rm e}^{ - i k t} u_k^{\rm in} (r) 
			\quad \Longrightarrow \quad 
			\left(  \frac{d^2}{d r ^2} + k^2  \right) u_k^{\rm in} (r) = 0 
			\quad 
			\Longrightarrow \quad  u_k^{\rm in} (r)  = N {\rm e}^{-i k r} \, , 
			\ee
			where $N$ is a normalisation that can be determined according to the scalar product:
			\be 
			\left(u_k^{\rm in} , u_{k'}^{\rm in} \right) = \delta_{k,k'} \, ,
			\ee			
			evaluated on the hyper-surface $\mathcal{I}^-$. 
			We ignore the outgoing $-ikr$ mode at $\mathcal{I}^-$ because 
			it propagates from infinity to infinity and is completely irrelevant.
			Finally, for the sake of simplicity we drop the ``0'' label, but we include the normalization factor. Therefore, taking of account of the definition (\ref{uvM}), the function $u_{k}(t,r)$ in (\ref{frt}) for the mode $k$ reads:
			\be 
			\label{inmode2} 
			u^{\rm in}_{k}(t,r) = \frac{1}{4 \pi \sqrt{k}} \frac{{\rm e}^{-i k v}}{r}  .
			\ee

			On the other hand, \underline{in the Schwarzschild spacetime (\ref{Schout})}, the d'Alembertian operator reads:
			\be 
			&& \left(- \frac{\partial^2}{\partial t ^2} +  \frac{\partial^2}{\partial r_*^2} - V_l (r)  \right)  v ^{\rm out}_{k,l}(t,r) +  \frac{R}{6} v ^{\rm out}_{k,l}(t,r)= 0 \, , 
			\quad 
			{\rm where} \quad
			V_l (r) = \left( 1 - \frac{2M}{r} \right) \left[  \frac{l(l+1) }{r^2} + \frac{2M}{r^3} \right] \, .
			\label{V0}
			\ee
			Since the Schwarzschild spacetime is Ricci flat, $R=0$. 
			According to the Hawking's original derivation, most of the  
			particle production happens at infinity and near the horizon where the potential vanishes, thus (\ref{V0}) simplifies to:
			\be 
			\left(- \frac{\partial^2}{\partial t ^2} +  \frac{\partial^2}{\partial r_*^2} \right)  v ^{\rm out}_{k}(t,r(r^*))  = 0 \, , 
			\label{KGS}
			\ee
			which in form is exactly like the wave equation in Minkowski spacetime (\ref{KGMin}). 
			
			Assuming the harmonic time dependence, i.e. 
			\be
			v ^{\rm out}_{k}(t,r)={\rm e}^{- i k t} v ^{\rm out}_{k}(r) \, , 
			\ee
			the equation (\ref{KGS}) simplifies to: 

			\be 
			\left( \frac{\partial^2}{\partial r_*^{2}} +k^2  \right) v ^{\rm out}_{k}(r(r_*)) = 0
			\quad 
			\Longrightarrow \quad  v_k^{\rm out} (r(r_*))  = N {\rm e}^{ + i k r_*} \, , 
			\ee

			Combining the radial and the time components according to (\ref{uvS}), finally, the solution reads:
			\be  
			v_k ^{\rm out} = \frac{1}{4 \pi \sqrt{k} } \frac{{\rm e}^{- i k u_{\rm out}}} {r} \, .
			\label{SolSch}
			\ee
			Having the solutions of the field equation in both the ``in'' and the ``out'' regions, we can proceed to compute the number of particles seen by an observer at $\mathcal I^+$.
			
			As we have shown in eq.(\ref{particNumb}), in order to get the particle number, we need to compute $|\beta_k|^2$, and 
				the Bogoliubov coefficient $\beta_{k k'}$ can be expressed in terms of the canonical inner product,
				\be  
				\label{beta3}
				\beta_{k k'}	= - (v_k ^{\rm out},  u_{k' }^{\rm in *}) = i \int_{\Sigma} d \Sigma  \, \,  n^\mu \left(v_k ^{\rm out} \partial_{\mu} u_{k' }^{\rm in}  - u_{k'} ^{\rm in} \partial _ \mu v_k ^{\rm out} \right),
				\ee
				where $n^\mu$ is a unit vector normal to the hypersurface $\Sigma$.
				Note that this integral is independent of the choice of the hypersurface $\Sigma$, as showed in \cite{Townsend}. Hence, in order to simplify the computation we evaluate the integral at $\mathcal I^-$, 
				\be \label{betascri}  \beta_{kk'}	= - (v_k ^{\rm out},  u_{k' }^{\rm in *}) = i \int_{\mathcal{I}^-} dv  \, r^2 d\Omega \left(v_k ^{\rm out} \partial_{v} u_{k' }^{\rm in}  - u_{k'} ^{\rm in} \partial _ v v_k ^{\rm out} \right).
				\ee   
				Since the integral is evaluated on one single Cauchy hypersurface, 
				we have to express both the ``in" and the ``out" modes in coordinates that are suitable for the specified hypersurface. 
				Hence, we express both $u_k^{\rm in}$ and $v_k ^{\rm out}$ in terms of the ``in" coordinates.

				The Minkowski and the Schwarzschild metrics in terms of the {\it{ingoing}} and {\it{outgoing}} coordinates are given in (\ref{Mink}) and (\ref{Schout}) respectively. 
				Therefore, in order to express the ``outgoing" coordinates in terms of the ``ingoing" ones, we need to perform the  matching of the Minkowski and the Schwarzschild metric at the null line $v = v_0$, which defines the boundary between the two regions. 
				
				According to (\ref{uvM}), the radial coordinate in the Minkowski space reads:
				\be
				r_{\rm in} = \frac{v - u_{\rm in}}{2}.
				\label{M2}
				\ee
				On the other hand according to (\ref{uvS}) in the Schwarzschild spacetime the tortoise radial coordinate reads:
				\be
				r_{* {\rm out}} = \frac{v - u_{\rm out}}{2}.
				\label{S2}
				\ee
				At the line $v=v_0$ (\ref{M2}), (\ref{S2}), and the tortoise coordinate $r_{*{\rm out} }$ are:
					\be
					&& r_{\rm in}(v_0, u_{\rm in}) = \frac{v_0 - u_{\rm in} }{2} \, , 
					\label{M3}\\
					&& r_{* {\rm out} }(v_0, u_{\rm out} )  = \frac{v_0 - u_{\rm out} }{2} 
					\label{S3}\, , \\
					&&  r_{* \rm{out}} = r_{\rm out} + 2 M \ln \frac{| r_{\rm out} - 2 M |}{2M} \, . 
					\label{Tor}
					\ee
					Replacing (\ref{Tor}) in (\ref{S3}) we get:
					\be
					r_{\rm in} + 2 M \ln \frac{| r_{\rm in} - 2 M |}{2M} = \frac{v_0 - u_{\rm out} }{2} 
					\label{S4}\, ,
					\ee
					where in (\ref{Tor}) we used the same radial coordinate for the Schwarzschild as well as the Minkowski spacetime, namely $r_{\rm out} = r_{\rm in} \equiv r$. The letter identification comes from the continuity of the angular part of the metric, namely, the metric of the two-sphere is the same in the two spacetime regions. 
					
					Finally, replacing (\ref{M3}) in (\ref{S4}) we end up with:
					\be
					&&  \frac{\cancel{v_0} - u_{\rm in} }{2} + 2 M \ln \frac{| r_{\rm in} - 2 M |}{2M} = \frac{\cancel{v_0} - u_{\rm out} }{2} 
					\label{S0}\\
					&& u_{\rm out} =   u_{\rm in}  - 4 M  \ln \Big|  \frac{ v_0  - u_{\rm in}  }{4 M } - 1 \Big|  	
					\label{S5}\, . 
					\ee
					Replacing (\ref{S5}) in (\ref{SolSch}), we get the solution $v_k ^{\rm out}$ in the ``in" coordinates, 
					\be  
					v_k ^{\rm out} = \frac{1}{4 \pi \sqrt{k} } \frac{{\rm e}^{- i k u_{\rm out}(u_{\rm in})}} {r} = 
					\frac{1}{4 \pi \sqrt{k} } \frac{{\rm e}^{- i k\left[    u_{\rm in}  - 4 M  \ln \Big|  \frac{ v_0  - u_{\rm in}  }{4 M } - 1 \Big|     \right] }}{r} 
					\, .
					\label{SolSch2}
					\ee
					In order to eliminate the singularity of the solution in $r=0$, we subtract to the solution (\ref{SolSch2}) the limit $v_k ^{\rm out}(r \rightarrow 0)$, namely we redefine the solution to be:
					\be  
					v_k ^{\rm out} =  
					\frac{1}{4 \pi \sqrt{k} } \frac{{\rm e}^{- i k\left[    u_{\rm in}  - 4 M  \ln \Big|  \frac{ v_0  - u_{\rm in}  }{4 M } - 1 \Big|     \right] }}{r} - \lim_{r \rightarrow 0} \frac{1}{4 \pi \sqrt{k} } \frac{{\rm e}^{- i k\left[    u_{\rm in}  - 4 M  \ln \Big|  \frac{ v_0  - u_{\rm in}  }{4 M } - 1 \Big|     \right] }}{r}
					\, .
					\label{SolSch3}
					\ee
					According to (\ref{uvM}), the limit $r \rightarrow 0$ is equivalent to the replacement $u_{\rm in} = v$. Therefore, the solution turns into:
					\be
					v_k ^{\rm out} =
					\frac{1}{4 \pi \sqrt{k} } \frac{{\rm e}^{- i k\left[    u_{\rm in}  - 4 M  \ln \Big|  \frac{ v_0  - u_{\rm in}  }{4 M } - 1 \Big|     \right] }}{r} -  \frac{1}{4 \pi \sqrt{k} } \frac{{\rm e}^{- i k\left[    v  - 4 M  \ln \Big|  \frac{ v_0  - v  }{4 M } - 1 \Big|     \right] }}{r}
					\, ,
					\label{SolSch4}
					\ee
					which satisfies $\lim_{r \rightarrow 0} v_k ^{\rm out} = 0$\footnote{Notice that (\ref{SolSch4}) is a solution of the d'Alembertian operator in Minkowski space (\ref{Mink}), but in coordinates $(u_{\rm in}, v)$ (\ref{uvM}), namely $\partial_{u_{\rm in}} \partial_v \phi = 0$.}. 
					Finally, we have to take into account that $u_{\rm out} \rightarrow + \infty$ for $u_{\rm in} = v = v_H \equiv v_0 - 4M$ (see \ref{S5}), which corresponds to $i^+$, thus  
					all the modes created for $v > v_H$ cannot reach $\mathcal{I}^+$. Therefore, (\ref{SolSch4}) has to be modified accordingly, 
					\be
					v_k ^{\rm out} =
					\frac{1}{4 \pi \sqrt{k} } \frac{{\rm e}^{- i k\left[    u_{\rm in}  - 4 M  \ln \Big|  \frac{ v_0  - u_{\rm in}  }{4 M } - 1 \Big|     \right] }}{r} -  \frac{1}{4 \pi \sqrt{k} } \frac{{\rm e}^{- i k\left[    v  - 4 M  \ln \Big|  \frac{ v_0  - v  }{4 M } - 1 \Big|     \right] }}{r} \Theta(v_H - v) 
					\, .
					\label{SolSch5}
					\ee

					\noindent Since we are interested in the expectation value of the number operator at  late time, we need accurate information about both the time and the energy of the excited modes. However, this is not achievable, as well known, and we have to turn the wave-modes into wave-packets \cite{Fabbri}.
					We coarse grain the mode $k$ and turn it into $j \epsilon$, where $\epsilon$ is a quantum of energy, and $j$ is the number of quanta. The value of $\epsilon$ determines the width of uncertainty over the energy. 
					Moreover, $\epsilon$ will act as a regulator in the equations that follow. 
					For instance, the $v^{\rm out} _{j n}$ modes read, 
					\be 
					v_{k} ^{\rm out} \,\, \rightarrow \,\,
					v_{j n} ^{\rm out} =
					-  \frac{1}{\sqrt \epsilon} \int _{j \epsilon} ^{(j+1) \epsilon} d k \, {\rm e} ^{2 \pi i k n/\epsilon }  v_k^{\rm out} \, ,
					\ee 
					which form a complete orthonormal set of modes for all values of $j$.
					The integer 
					$n$ can be interpreted as time, as we discretize the coordinate 
					$u_{\rm out} = 2 \pi n/\epsilon$, and for $n \to \infty$, $u_{jn}^{\rm out}$ represent the mode at the future null-infinity.
					Now we are ready to compute the $\beta_{j n \, k'}$ integral at $\mathcal I^-$ because we have both the ``in" and the ``out" modes expressed in terms of the coordinates at $\mathcal I^-$ (``in").  
					With a simple integration by parts of  (\ref{betascri}) and assuming the boundary terms to vanish, (\ref{betascri}) turns into:  
					\be \beta_{ j n \, k'}  = 2 i \int_{\mathcal I ^-} dv r^2 d \Omega  \, v_{j n} ^{\rm out} \, \partial_v u_{k'} ^{\rm in} 
					= \frac{-1}{2 \pi \sqrt \epsilon}  \int_{- \infty} ^{v_H}  dv \int _{j \epsilon} ^{(j+1) \epsilon} dk \, {\rm e}^{2 i \pi k n/\epsilon } v_k^{\rm out}  \partial _v u_{k'}^{\rm in}
					.
					\label{Dami}
					\ee

					\noindent Replacing the expressions for the wave modes (\ref{inmode2}) and (\ref{SolSch5}) in (\ref{Dami}),
					\be  \label{betabvc}
					\beta_{ j n \, k'} = - \frac{1}{2 \pi \sqrt \epsilon} \int_{- \infty} ^{v_H}  dv   \int _{j \epsilon} ^{(j+1) \epsilon} dk \, {\rm e}^{2 i \pi k n/\epsilon }  \sqrt{ \frac{k'}{k} } {\rm e}^{- i k( v_\text{H}- 4 M \ln \frac{v_{\rm H} - v}{ 4 M} )  - i k' v} \, .
					\ee
					Performing the integral over $k$ and taking into account that $k$ varies over a very small interval, 
					\be 
					\beta_{ j n \, k'} =  \frac{-{\rm e}^{- i (k_j + k') v_{\rm H} }  }{ \pi \sqrt \epsilon} \sqrt{ \frac{k'}{k_j} } \int_0^\infty dx   \, {\rm e}^{ i  k' x} \sin \frac{\epsilon L/2} {L}     {\rm e}^{ i L k_j } \, ,
					\ee
					where we made the change of variables: 
					$x \equiv v_{\rm H}  - v$, $L\equiv \frac{2 \pi n}{\epsilon} + 4 M \ln \frac{x}{4M} $, and  $k_j \equiv j \epsilon \sim  (j+\frac{1}{2}) \epsilon $.

					\be 
					\beta_{ j n \, k'} =  \frac{-{\rm e}^{- i (k_j + k') v_{\rm H} }  }{ \pi \sqrt \epsilon} \sqrt{ \frac{k'}{k_j} } \int_{- \infty} ^{v_H}  dv   \int _{j \epsilon} ^{(j+1) \epsilon} dk \, {\rm e}^{ i  k' x} \sin \frac{\epsilon L/2} {L}     {\rm e}^{ i L k_j } \, .
					\label{betan}
					\ee
					Following exactly the same steps, one finds $\alpha_{k k'}$:  
					\be \alpha_{kk'} = ( v_k ^{\rm out}, u_{k'} ^{\rm in})
					= - 2i \int _ {\mathcal I^-}  dv r^2 d \Omega v_k ^{\rm out}  \partial_v u_{k' } ^{\rm in *}   = \frac{-{\rm e}^{- i (k_j - k') v_{\rm H} }  }{ \pi \sqrt \epsilon} \sqrt{ \frac{k'}{k_j} }  \int_0^\infty dx   \, {\rm e}^{- i  k' x} \sin \frac{\epsilon L/2} {L}     {\rm e}^{ i L k_j } \, .
					\label{alphan}
					\ee
					Let us define the integral involved in (\ref{betan}) and (\ref{alphan}) by:
					\begin{eqnarray}
						&& I(k') = \int_0^\infty dx   \, {\rm e}^{ i  k' x} \sin \frac{\epsilon L/2} {L}  {\rm e}^{ i L k_j } \, , \\
						&& I(-k')  = \int_0^\infty dx   \, {\rm e}^ {-i  k' x} \sin \frac{\epsilon L/2} {L}     {\rm e}^{ i L k_j }.
					\end{eqnarray}
					In order to evaluate the above integrals we move to the complex plane $x$ and we replace $x=iy$. Indeed, the integral along the imaginary axis is equivalent to the one along the real axis because the integrand has no poles. Hence: 
					\begin{eqnarray}
						I(k') = - i \int_0^{+\infty}  dy \, {\rm e}^{-k' y} \sin \frac{\epsilon L_y/2} {L_y}     {\rm e}^{ i L_y k_j } ,
					\end{eqnarray}
					where $L_y = \frac{2 \pi n}{\epsilon} - 4M \ln \frac{iy}{4M}$ and we replaced $\ln [ -i f]= -i\frac{\pi}{2} + \ln{f}$. Finally, 
					\be 
					I(k') = - i {\rm e}^{2 M \pi k_j}  {\rm e }^{2 \pi i n k_j/\epsilon} \int_0^{+\infty}  dy \, {\rm e}^{-k' y} \sin \frac{\epsilon L_y/2} {L_y}     {\rm e}^{ i 4M k_j \ln(\frac{y}{4M})} .
					\ee
					Similarly, in $I(-k')$ we replace $x\to i z$, to end up with:
					\be 
					I(-k') = i {\rm e}^{-2 M \pi k_j}  {\rm e }^{2 \pi i n k_j/\epsilon} \int_0^{+\infty}  dz \, {\rm e}^{k' z} \sin \frac{\epsilon L_z/2} {L_z}     {\rm e}^{ i 4M k_j \ln(\frac{z}{4M})} .
					\ee
					Where $L_z\equiv \frac{2 \pi n }{\epsilon} + 4Mi\frac{\pi}{2} + 4M \ln \frac{z}{4M}$.
					
					Finally, we notice that for an extremely narrow wave-packet, $\epsilon \ll 1 $, $\epsilon L_z\sim \epsilon L_y$ at very late times $\frac{n}{\epsilon}\to \infty$,
					\be 
					I(k')= - {\rm e}^{4 \pi M k_j } I(-k'),
					\ee
					leading to the following relation between the Bogoliubov coefficients, 
					\be 
					\alpha_{jn, k'}= - {\rm e}^{4 \pi M k_j }  {\rm e}^{2 i  k' v_{\rm H} }\beta_{jn,k'}.
					\ee
					Renaming $nj$ by $k$ and taking the absolute value, the above expression boils down to:

						\be 	
						|\alpha_{k k'} | = e^{4 \pi M k} | \beta_{k k'} | \, , 
						\ee 	
						which combined with the normalization condition for $\alpha$ and $\beta$, namely
						\be \int ^{+ \infty} _0 d k' ( | \alpha_{k k' } |^2 - | \beta_{k k' } |^2 ) = 1 ,
						\ee
						gives the number of particles 
						\be \langle  N_k ^{\rm out} \rangle  = \int^{+ \infty} _ 0 dk' |\beta_{k k' } |^2 = \frac{1}{ {\rm e}^{8 \pi M k} - 1 } . 	
						\ee
					}

\end{document}